\documentclass[12pt]{article}

\usepackage{amsmath,amssymb,amsthm,cite,enumitem,mathtools,xcolor}
\usepackage[margin=3cm]{geometry}

\usepackage[T1]{fontenc}
\usepackage[osf]{newtxtext} 
\usepackage[nosymbolsc,cmintegrals]{newtxmath}
\usepackage[cal=euler,frak=euler]{mathalfa} 
\usepackage{bm} 

\title{Ensuring locality in QFT via string-local fields}

\author{José M. Gracia-Bondía$^{1,2}$ and Joseph C.~Várilly$^3$
\\ [12pt]
{\footnotesize $^1$ Departamento de Física Teórica,
Universidad de Zaragoza, Zaragoza 50009, Spain}\\[3pt]
{\footnotesize $^2$ Escuela de Física,
Universidad de Costa Rica, San Pedro 11501, Costa Rica}\\[3pt]
{\footnotesize $^3$ Escuela de Matemática,
Universidad de Costa Rica, San José 11501, Costa Rica}\\[3pt]
}

\date{7 July 2023}

\DeclareMathOperator{\T}{T}         
\DeclareMathOperator{\tsum}{{\textstyle\sum}} 

\newcommand{\al}{\alpha}            
\newcommand{\bt}{\beta}             
\newcommand{\dl}{\delta}            
\newcommand{\eps}{\varepsilon}      
\newcommand{\Ga}{\Gamma}            
\newcommand{\ga}{\gamma}            
\newcommand{\ka}{\kappa}            
\newcommand{\La}{\Lambda}           
\newcommand{\la}{\lambda}           
\newcommand{\Om}{\Omega}            
\newcommand{\sg}{\sigma}            
\newcommand{\Th}{\Theta}            
\newcommand{\vf}{\varphi}           

\newcommand{\kaa}{{\underline{\kappa}}} 
\newcommand{\muu}{{\underline{\mu}}}    
\newcommand{\nuu}{{\underline{\nu}}}    

\newcommand{\bD}{\mathbb{D}}        
\newcommand{\bR}{\mathbb{R}}        
\newcommand{\bS}{\mathbb{S}}        

\newcommand{\sD}{\mathcal{D}}       

\newcommand{\pt}{{\mathrm{p}}}      
\newcommand{\sym}{{\mathrm{sym}}}   
\newcommand{\tree}{{\mathrm{tree}}} 

\bmdefine{\CC}{C}             
\bmdefine{\ddel}{\partial}    
\bmdefine{\ee}{e}             
\bmdefine{\KK}{K}             
\bmdefine{\lll}{l}            
\bmdefine{\LL}{L}             
\bmdefine{\lll}{l}            
\bmdefine{\mm}{m}           
\bmdefine{\nn}{n}           
\bmdefine{\PP}{P}           
\bmdefine{\pp}{p}           
\bmdefine{\rr}{r}           
\bmdefine{\ssg}{\sigma}     
\bmdefine{\SSS}{S}          
\bmdefine{\ttt}{t}          
\bmdefine{\ttau}{\tau}      
\bmdefine{\tth}{\theta}     
\bmdefine{\uu}{u}           
\bmdefine{\vv}{v}           
\bmdefine{\ww}{w}           
\bmdefine{\VV}{V}           
\bmdefine{\xii}{\xi}        
\bmdefine{\xx}{x}           
\bmdefine{\yy}{y}           
\bmdefine{\zz}{z}           
\bmdefine{\zero}{0}         
\bmdefine{\zzeta}{\zeta}    

\newcommand{\del}{\partial}         
\newcommand{\downto}{\downarrow}    
\newcommand{\otto}{\leftrightarrow} 
\newcommand{\ovl}{\overline}        
\newcommand{\x}{\times}             
\newcommand{\0}{{\vphantom{0}}}     
\newcommand{\7}{\dagger}            
\newcommand{\8}{\bullet}            

\newcommand{\delotto}{\overset{\leftrightarrow}{\partial}\!} 

\newcommand{\sump}{\sideset{}{'}\sum} 

\newcommand{\half}{{\mathchoice{\thalf}{\thalf}{\shalf}{\shalf}}}
\newcommand{\shalf}{{\scriptstyle\frac{1}{2}}} 
\newcommand{\thalf}{\tfrac{1}{2}}   
\newcommand{\quarter}{\tfrac{1}{4}} 

\newcommand{\set}[1]{\{\,#1\,\}}  
\newcommand{\vev}[1]{\langle\!\langle#1\rangle\!\rangle} 
\newcommand{\word}[1]{\quad\text{#1}\quad} 

\def\wick:#1:{\,\mathopen:#1\mathclose:\,} 

\def\epsi^#1_#2{\eps^{#1}{}_{\!#2}} 
\def\epsii_#1^#2{\eps_{#1}{}^{\!#2}} 
\def\lLa^#1_#2{\Lambda^{#1}{}_{#2}}  

\newcommand{\braket}[2]{\langle#1\mathbin|#2\rangle} 

\def\duo<#1,#2>{\langle#1,#2\rangle} 

\theoremstyle{plain}

\makeatletter
\renewcommand{\section}{\@startsection{section}{1}{\z@}%
                       {-3.5ex \@plus -1ex \@minus -.2ex}%
                       {2.3ex \@plus.2ex}%
                       {\normalfont\large\bfseries}}
\renewcommand{\subsection}{\@startsection{subsection}{2}{\z@}%
                       {-3.25ex \@plus -1ex \@minus -.2ex}%
                       {1.5ex \@plus .2ex}%
                       {\normalfont\normalsize\bfseries}}
\makeatother

\begin{document}

\maketitle

\begin{abstract}
String-local fields constitute a relatively new tool for solving
quantum field theory, stressing and embodying locality and positivity.
We examine here their usefulness -- as well as some drawbacks.
Starting from \textit{just} the physical masses and charges of the
known particles, and bringing string independence in the
Bogoliubov--Epstein--Glaser (BEG) theory framework
\cite{BogoliubovS80, EpsteinG73, Duetsch18}, regarded \textit{as a
means of discovery}, one finds the allowed couplings of quantum fields
associated to those particles, and thereby recovers all of the
Standard Model (SM) without invoking theoretical prejudices. One of
the outcomes is the requirement that the fields be governed by
reductive Lie algebras. Another is the need for at least one scalar
particle. Yet another is chirality of interactions mediated by massive
particles. There is no room in this formulation for ``global'' gauge
invariance as an \textit{a~priori} construct.

Armed with this modern weapon, we reassess here a few classical and
recent conundra. In particular, we examine new perspectives in
Cosmology from the adoption of string-local fields.
\end{abstract}

\medskip 

\begin{flushright}
\begin{minipage}[t]{22pc}
\small \itshape
Science flourishes best when it uses freely all the tools at hand
[\dots] A new tool always leads to new and unexpected discoveries,
because Nature's imagination is richer than ours.
\par \smallskip \hfill \upshape
--- Freeman Dyson \cite{Dyson95}
\end{minipage}
\end{flushright}

\medskip

\textit{Keywords}:
Locality / Positivity / Epstein--Glaser theory / String independence

\newpage 

\tableofcontents

\section{Introduction}
\label{sec:lucet}

\subsection{The scenario}
\label{sec:omnibus}

Modern algebraic quantum field theory (AQFT), with its emphasis on
algebras of observables rather than on fields \cite{Yngvason15,
Witten18}, and the customary theoretical constructions and procedures
to deal with the Standard Model (SM) phenomena, in particular the
model's reliance on classical Lagrangians and gauge theory, have been
sociologically divorced for many years. Since it is hard to dispute
that QFT should stand on its own legs, ``not on classical crutches''
\cite{Jordan29}, a revision of this sorry state of affairs is
permanently on the charts; and indeed such is the purpose of several
relatively recent papers which attempt to formulate perturbative gauge
theories in terms of $C^*$-algebras. See in this respect
\cite{BrunettiDFR21} and the antecedents cited there.

In this paper we take a different tack. Consider the Wigner
classification of elementary particles as unirreps of the restricted
Poincaré group~\cite{Wigner39}. It establishes three types:
\begin{enumerate}
\item 
Massive $(m > 0)$ particles.
\item 
Massless $(m = 0)$ particles of finite helicity $h$.
\item 
Massless $(m = 0)$ particles of unbounded helicity.
\end{enumerate}

Each of these types can be represented in infinitely many ways by
quantum fields \cite[Chap.~5]{Weinberg95I}. Of course, some of the
latter are more useful and/or natural than others -- and most work on
QFT is done in terms of \textit{potential fields}. On the other hand,
one would expect the descriptions to ensure \textit{positivity}
(leading to a probabilistic interpretation of the formalism),
\textit{covariance}, and \textit{causal localization}.

\medskip

As is well known, the picture in terms of the fields is rather more
complicated:
\begin{enumerate}
\item 
All massive particles can be described by potential fields that obey
positivity, covariance and causal localization.%
\footnote{This is not quite obvious, but can be seen to hold --
consult \cite[Ch.~8]{Schwartz14} for the simplest case.}
Their high-energy behaviour grows steadily worse with spin.

\item 
Massless particles with $h = 0$ or $|h| = \half$ enjoy the same above
properties as do the corresponding massive particles. Between them and
their massive counterparts, there is a smooth transition of
correlation functions as $m \downto 0$.

\item 
Massless particles of helicity $|h| \geq 1$ have field descriptions
that possess those properties as well; however, the apparently
``natural'' ones do not. A paradigmatic example is the description of
photons: the Maxwell field $F^{\mu\nu}(x)$ is positive, covariant and
local, but the electrodynamic potential $A^\nu(x)$ is neither.
Similarly for gluodynamics. Since violation of positivity conflicts
with the probabilistic interpretation of quantum mechanics, to salvage
positivity for \textit{observables}, one is forced to introduce ghost
fields and the like.

Moreover, the current theory contains some counter-intuitive negative
results, such as the Weinberg--Witten theorem, stating that for
$|h| \geq 2$ no stress-energy-momemtum can exist whose zero-components
generate the Poincaré momenta. Another somewhat strange result is the
discontinuity of scattering amplitudes in the mass $m$ at $m = 0$ and
$|h| = 2$ (the Zakharov--van~Dam--Veltman gap) \cite{Zakharov70,
VanDamV70}.

\item 
On the face of it, fields describing massless particles of unbounded
helicity enjoy neither positivity nor causal localization. Long ago
Yngvason~\cite{Yngvason70} proved that old algebraic QFT, based on the
Wightman axioms for the fields, does not encompass them.

\end{enumerate}

Given this picture, the received wisdom in the SM, that quantum
particles are naturally massless and that their masses are ascribable
to ``spontaneous symmetry breaking'' of a ``gauge symmetry'', is
somewhat suspect. The latter improbable construct is obviously the
outcome of trying to fit the scalar particle(s) in the Procrustean bed
of gauge theory. But gauge \textit{invariance}, rather than a physical
symmetry, is a formal instrument to extract a physical sub-theory from
an unphysical formalism. Nowadays this is recognized even in
textbooks: ``Gauge invariance is not physical and is not a symmetry of
nature \dots\ [It] is merely a redundancy of description we introduce
to be able to describe the theory with a local Lagrangian''
\cite[Sect.~8.6]{Schwartz14}. It is hard to see in which sense it can
be ``broken'' -- consult \cite[App.~C]{Aurora}. We shall return to the
matter.

\medskip

The situation as regards the relation between the classes numbered
1~and~3 in the second list above \textit{and} of the handling of
class~4 has drastically changed in relatively recent times, thanks to
the emergence of rigorous \textit{string-localized field theory}. Now,
string-localized quantum fields (SLFs) are definitely non-exotic. In a
heuristic version recognizably similar to the modern one, they were a
brainchild of Mandelstam~\cite{Mandelstam62}. In a different vein,
their pertinence follows from work by Buchholz and
Fredenhagen~\cite{BuchholzF82} in the eighties, on the construction of
particle states in AQFT. At about the same time, SLFs were being
further developed by Steinmann \cite{Steinmann82, Steinmann84,
Steinmann85}. The papers by Jordan~\cite{Jordan35} and
Dirac~\cite{Dirac55} are rightly deemed to be precedents. All these
works strive to break free from the shortcomings of the
$A^\nu(x)$-type fields.

The new development was made possible by the advent of \textbf{modular
localization} and the Bisognano--Wichmann theorem, which most
effectively marry geometric with analytic aspects of QFT. For a clear
introduction to modular localization, consult~\cite{Yngvason15}. Key
articles on what concerns us here were \cite{Mund01, BrunettiGL02,
Mund07}. There it is proved that quantum fields connecting the vacuum
with the single-particle states, associated to any kind of Wigner
particle, can be localized on spacelike cones in Minkowski space. It
turns out that the half-line limits of such cones (the strings) are
satisfactory for most purposes. We shall denote by~$H$ the open set of
spacelike vectors that make up the strings.%
\footnote{We employ the mostly-negative Minkowski metric.}
In the subsequent period Yngvason, Mund and Schroer \cite{MundSY04,
MundSY06} were able to construct SLFs, localized in such spacelike
half-lines, associated with all irreducible Wigner representations of
the Poincaré group, including the particles of our class~3.

\medskip

Following \cite{GassDiss22}, we summarize a few main traits,
advantages, and uses of SLFs.
\begin{enumerate}

\item 
Their improved scaling behaviour. It so happens that the high-energy
behaviour of SLFs is always the one corresponding to either $s,h = 0$
or $s,|h| = \half$, \textit{irrespective} of their helicity or spin.

\item 
SL fields satisfy the Bisognano--Wichmann property, relating the
modular group associated with a spatial wedge in momentum space to
corresponding Lorentz boosts.

\item 
Smooth transition of the correlation functions as $m \downto 0$ is
kept, irrespectively of spin. How this works may be illustrated in the
case of massive and massless QED. The coupling to the indefinite
positive Maxwell potential $A^F$ ($F$~for ``Feynman gauge'') is
replaced by a coupling $j^\mu A^\pt_\mu$ to the massive Proca
potential $A^\pt$. This avoids negative-norm states, but the
interaction is apparently non-renormalizable because of the UV
dimension~$2$ of the Proca potential. Now a decomposition
$A^\pt_\mu(x) = A_\mu(x,e) - m^{-1} \del_\mu a(x,e)$ into a
string-localized potential and its \textit{escort field}~$a$ is
brought to bear: $A_\mu(x,e)$ has UV dimension~$1$ and is regular at
$m = 0$. The UV-divergent part of the interaction is carried away by
the escort field:
$-m^{-1} j^\mu \del_\mu a(e) = - \del_\mu(m^{-1} j^\mu a(e))$ is a
total derivative and is discarded from the coupling. The remaining
string-localized interaction $j^\mu A_\mu(x,e)$ is equivalent to the
point-localized one and keeps the good ultraviolet behaviour at
$m = 0$.

\item 
SLF theory is formulated exclusively with physical degrees of freedom.
Consequently, the concept of ``gauge invariance'' as a fundamental
principle is absent. A formal similarity to geometric connection
theory and classical Lagrangians of the allowed couplings for SLFs is
all that remains. We show in the body of the paper how this overdue
banishing of unphysical degrees of freedom comes about.

\end{enumerate}

Before continuing, we call to the reader's attention the paper by
Herdegen~\cite{Herdegen22}, where SL fields for photons are compared
to radial and almost-radial gauges' formalism for electrodynamics.
There the suggestion is made of using full spacelike lines instead of
half-lines.

In general, SLFs offer few calculational advantages. Their role is
more of a detective nature: wherever in QFT, and in the SM in
particular, one ``smells a rat'', they often come in handy. The upshot
is that large no-go territories for quantum field theory are now
trespassed. A few examples are:
\begin{itemize}

\item
Chirality of the electroweak sector of the SM is \textit{derived} 
without special assumptions.

\item
All of gluodynamics is established without previous assumptions.

\item
The need for a scalar particle -- say, the Higgs' particle -- in
the~SM is established without recourse to unobservable mechanisms.

\item
The separation of helicities in the massless limit of higher spin
fields is clarified. The van Dam-Veltman--Zakharov discontinuity
\cite{Zakharov70, VanDamV70} at the $m \downto 0$ limit of massive
gravitons is resolved.

\item
Unimpeachable stress–energy-momentum (SEM) tensors for massless fields
of \textit{any} helicity are constructed \cite{MundRS17a, MundRS17b}
-- allowing for gravitational interaction, and in particular flouting
the Weinberg--Witten theorems~\cite{WeinbergW80}.

\item
The strong $\bm{CP}$ (pseudo)problem is exorcised.

\end{itemize}

\subsection{Dispensing with ``renormalization''}
\label{ssc:no-coffee-break}

A recent conceptual step forward, related to the issues of this paper,
can be encountered in~\cite{MooijSh23}. Its authors argue that
(perturbative) QFT ought to be defined by ``divergencies-free''
methods, resulting in particular that the fashionable ``hierarchy
problem'' is exorcised, all postulated fine-tunings in theories with
well-separated energy scales being discarded.

The main tool in \cite{MooijSh23} is renormalization by
Callan--Symanzik-type differential equations. We hold, and set out to
show in this paper, that the BEG framework can be used to the same
purpose, and generally as a means of discovery. In so doing, we
contend that to regard this framework as a mere tool of
renormalization is somewhat misleading.%
\footnote{The present authors are not entirely guiltless in this
regard.}
Indeed, Epstein and Glaser originally \cite{EpsteinG73} stressed not
so much ``renormalization'', but rather the pursuit of
\textit{locality} in perturbative quantum field theory. Like
Scharf~\cite{Scharf16}, in most instances we shall choose to speak of
\textit{normalization} rather than of ``\textit{re}normalization'',
inasmuch as no manipulation of infinities takes place in an
Epstein--Glaser approach.

\subsection{Aims and plan of the paper}
\label{ssc:what-we-want}

In the spirit of algebraic quantum field theory, one contemplates here
the particles regarded as elementary in the SM as described simply by
their \textit{masses} (couplings to the $h = \pm 2$ particle) and
\textit{charges} (couplings to charged $h = \pm 1$ particles), as
determined by the experiments. \textit{All} the details of the SM
interactions are to be retrieved on this basis. Now, string-localized
quantum fields come into the world with a powerful heuristic
principle: the string ``ought not to be seen''. That is to say, the
construction of physical observables and scattering amplitudes cannot
depend on the string coordinates. The principle will be used in the
BEG framework for scattering theory. Coming to the SM, we consider
here essentially full \textit{flavourdynamics} and
\textit{gluodynamics} (rather than full QCD) in turn. Gleaning from
previous work of ours on these subjects~\cite{Rosalind, Borisov}, all
the main traits of the SM, including the presence of at least one
scalar particle, are recovered in a constructive, \textit{bottom-up}
approach.

In Appendix~A we describe the structure of propagators for SL fields.

In Appendix~B we ponder the so-called ``strong~CP problem''.

In Appendix~C, in a more speculative vein, we deal with another
subversive aspect of SLFs, as regards higher-spin fields.

\section{On the relation between potentials and field strengths}
\label{sec:sorte}

For a generic free quantum field on Minkowski space we may write
$$
\Psi_l(x) = \sum_\sg \int d\mu_m(p)\, 
\bigl[ u_l(p,\sg) a(p,\sg) e^{-i(px)}
+ v_l(p,\sg) a^\7(p,\sg) e^{i(px)} \bigr].
$$
We would like this to transform as 
\begin{equation}
U(\La) \Psi_l(x) U(\La)^\7 
= \sum_{\bar l} D_{l\bar l}(\La^{-1}) \Psi_{\bar l}(\La x),
\label{eq:pushed-around} 
\end{equation}
with $U(\La)$ denoting the second quantization of the pertinent
unirrep of the Lorentz group, and $D$ being some matrix representation
of that group. Let $d$ be a~unirrep for a little group, say for~$p_0$.
There are $(\dim D \x \dim d)$ coefficient matrices $u(p)$, $v(p)$ for
$p$ in the mass-$m$ hyperboloid, such that
$u(\La p)\,d(R(\La, p)) = D(\La)\,u(p)$;
$v(\La p)\,\ovl{d(R(\La, p))} = D(\La)\,v(p)$, where $R(\La,p)$ is the
``Wigner rotation'' in the little group. For a complex scalar field:
$$
\vf_m(x) = \int d\mu_m(p)\, 
\bigl[ c(p) e^{-i(px)} + d^\7(p) e^{i(px)} \bigr],
$$
and these relations trivially and Eq.~\eqref{eq:pushed-around} hold.
Is same true of the standard~potential
$$
A_\mu(p) = \sum_{h=\pm 1} \int d\mu(p)\, 
\bigl[ u^h_\mu(p) a_h(p) e^{-i(px)} 
+ \ovl{u^h_\mu(p)} a_h^\7(p) e^{i(px)} \bigr],
$$
where $h$ denotes helicity? No, it is not. For a messy linear
combination $\Ga$ of creation and annihilation operators:
$$
U(\La) A_\mu(p) U(\La)^\7 
= \lLa^\nu_\mu A_\nu(\La x) + \del\Ga(x,\La).
$$
No intertwiners $u,v$ avoiding this can be found.

\goodbreak 

Instead, with the help of polarization vectors $\eps^h_\nu$ (zweibeins
or dreibeins), one may construct the Faraday covariant tensor fields
on Hilbert space,
\begin{equation}
F_{\mu\nu}(x) = \sum \int d\mu(p)\, 
\bigl[ u^h_{\mu\nu}(p) a_h(p) e^{-i(px)}
+ \ovl{u^h_{\mu\nu}(p)} a_h^\7(p) e^{i(px)} \bigr],
\label{eq:Faraday} 
\end{equation}
with $u^h_{\mu\nu}(p) = -i p^\0_{[\mu} \eps^h_{\nu]}(p)$, where $\sum$
denotes $\sum_{h=\pm 1}$ in the massless case and
$\sum_{-1\leq h\leq 1}$ in the massive case. Then, as is well known,
$$
U(\La) F_{\mu\nu}(p) U(\La)^\7 
= \lLa^\rho_\mu \lLa^\sg_\nu F_{\rho\sg}(\La x).
$$

Consider now semi-infinite strings $S_{x,e} := x + \bR^+ e$, where $e$
is a spacelike $4$-vector, and
$$
A_\mu(x,e) := I_e F_{\mu\nu}(x) e^\nu 
\equiv \int_0^\infty ds\, e^\nu F_{\mu\nu}(x + se).
$$
Clearly $I_e F_{\mu\nu} = I_{\la e} F_{\mu\nu}$ for $\la > 0$. We
understand
$$
\int_0^\infty dt\, e^{\pm itx} 
= \lim_{\eps\downto 0} \frac{\pm i}{x \pm i\eps}
=: \frac{\pm i}{x \pm i0}\,.
$$
Then
\begin{align}
A_\mu(x,e) &= \sum_{h=\pm1} \int d\mu(p)\, 
\bigl[ u^h_\mu(p,e) a_h(p) e^{-i(px)}
+ \ovl{u^h_\mu(p,e)} a_h^\7(p) e^{i(px)} \bigr],
\notag \\
\text{with } u^h_\mu(p,e) 
&:= \frac{u^h_{\mu\nu}(p,e) e^\nu}{i[(pe) - i0]} 
= \eps^h_\mu(p) - \frac{(\eps^h(p) e)}{(pe) - i0}\,p_\mu.
\label{eq:great-potential} 
\end{align}
These $u$ are \textit{bona fide} intertwiners:
$$
u_h(\La p, \La e)\,d(R(\La,p)) = D(\La)\,u_h(p,e),
$$
where $R(\La,p)$ is the ``Wigner rotation''.

It is often helpful to invert these formulas in order to obtain the
Fock space annihilation and creation operators. Using
$\bigl( u^h_\mu(p,e)\,\ovl{u^{h'}_\mu(p,e')} \bigr) 
= \bigl( \eps_h(p)\,\bar\eps_{h'}(p) \bigr) = -\dl_{hh'}$, one finds:
\begin{align*}
a_h(p) &= -i \int_{x^0=t} d^3x\, 
\ovl{u^h_\mu(p,e)} e^{-i(px)}\, \delotto_0\, A^\mu(x,e)
\\
a_h^\7(p) &= \int_{x^0=t} d^3x\, 
u^h_\mu(p,e) e^{i(px)}\, \delotto_0\, A^\mu(x,e).
\end{align*}

For arbitrary mass and integer spin or helicities, denote the
associated field strengths (symmetric under any exchange of pairs
$(\mu_j,\nu_j) \otto (\mu_k,\nu_k)$) by 
$F_{[\mu_1,\nu_1]\cdots [\mu_s,\nu_s]}(x)$, often shortened to
$F_{[\muu,\nuu]}(x)$. The corresponding string-localized
\textit{potential}, symmetric under exchange of its indices, is
defined by line integrals in the direction~$e$:
\begin{align}
A_{\mu_1\cdots\mu_s}(x,e) 
&:= \int_0^\infty \!\cdots\! \int_0^\infty dt_1\cdots dt_s\,
F_{[\mu_1,\nu_1]\cdots [\mu_s,\nu_s]} \biggl(
x + \sum_{i=1}^s t_i e \biggr) e^{\nu_1} \cdots e^{\nu_s}
\nonumber \\
&=: I_e^s F_{[\mu_1,\nu_1]\cdots [\mu_s,\nu_s]}(x)
e^{\nu_1} \cdots e^{\nu_s}.
\label{eq:high-potential} 
\end{align}
Using $(e\,\del_x) I_e = -1$ as well as the Bianchi identities for the
field strength, one can check that
$A_\muu(x,e) \equiv A_{\mu_1\cdots\mu_s}(x,e)$ is indeed a potential
for $F_{[\muu,\nuu]}(x)$ \cite{MundDO17}. Moreover, $A_\muu(x,e)$ is
\textit{axial} with respect to the string variable:
$e^{\mu_1} A_{\mu_1\cdots\mu_s}(x,e) = 0$. Furthermore, in the
massless case $\del^{\mu_1} A_{\mu_1\cdots\mu_s}(x,e) = 0$.

By their definitions, in both the massless and massive cases, the
string-localized $A_\muu$ live on the \textit{same} Hilbert space as
the field strengths. \textit{Positivity, locality and Poincaré
invariance} hold:
\begin{gather*}
[A_{\mu_1\mu_2}(x,e), A_{\mu_1\mu_2}(x',e')] = 0 
\word{if} \bigl( x + \bR_+ e - x' - \bR_+ e' \bigr)^2 < 0;
\\
U(a,\La)\, A_{\mu_1\mu_2}(x,e)\, U^\7(a,\La) 
= \lLa^{\ka_1}_{\mu_1} \lLa^{\ka_2}_{\mu_2} 
A_{\ka_1\ka_2} \bigl( \La x + a, \La e \bigr);
\end{gather*}
where $U(a,\La)$ denotes, as before, the second quantization of the
corresponding unirrep of the Poincaré group, and we have taken $s = 2$
or $|h| = 2$ to simplify notations. Similarly with anticommutators for
fermionic string-localized fields.

\medskip

For practical purposes, we hasten to introduce translation-invariant
\textit{two-point functions} for SLFs. Given two string-localized
fields of the same mass $m \geq 0$, we define
\begin{align*}
\vev{X(x,e)\,X'(x',e')} 
&:= \braket{\Om}{X(x,e)\,X'(x',e')\,\Om}
\\
&= \int\frac{d^4p}{(2\pi)^3}\, \dl(p^2 - m^2)\, \theta(p^0)\, 
e^{-i(p(x-x'))} M_m^{XX'}(p,e,e'),
\end{align*}
where $\Om$ denotes the vacuum state and $M_m^{XX'}$ or simply
$M^{XX'}$ will be a polynomial in~$p$. For the electromagnetic field
and its massive look-alike, by use of the intertwiners:
\begin{equation}
M^{FF}_{\mu\nu\rho\sg}(p) = g_{\mu\sg} p_\nu p_\rho
+ g_{\nu\sg} p_\mu p_\rho - g_{\mu\rho} p_\nu p_\sg
- g_{\nu\sg} p_\mu p_\rho.
\label{eq:field-of-play} 
\end{equation}
Also
\begin{align}
- M^{AA}_{\mu\rho}(p,e,e') 
&= g_{\mu\rho} - \frac{e_\rho p_\mu}{(pe) - i\eps}
- \frac{e'_\mu p_\rho}{(pe') + i\eps}
+ \frac{(ee') p_\mu p_\rho}{[(pe) - i\eps]\,[(pe') + i\eps]}
\notag \\
&=: E_{\mu\rho}(p,e,e').
\label{eq:potential-player} 
\end{align}

Formulas like \eqref{eq:high-potential} will require translation into 
momentum space. With some abuse of notation:
\begin{align}
I_e f(p) &:= \int d^4x \int_0^\infty ds\, e^{i(px)} f(x + se)
= \int d^4x \int_0^\infty ds\, e^{i((p - se)x)} f(x)
\nonumber \\
&= f(p) \int_0^\infty ds\, e^{-is(pe)}
=: -\frac{if(p)}{(pe) - i\eps}\,.
\label{eq:untune-that-string} 
\end{align}

\subsection{Escort fields and the $m \downto 0$ process}
\label{ssc:walk-with-me}

As indicated above, in the case of a massive spin-$1$ field, the Proca
potential can be retrieved as
$A^\pt_\nu(x) := A_\nu(x,e) - m^{-1} \del_\nu a(x,e)$, where the
\textit{escort field} $a(x,e)$ ``carries away'' the UV divergence of
$A^\pt_\nu(x)$ as $m \downto 0$ since it contributes a total
divergence to the coupling $j^\nu A^\pt_\nu$. This escort, a scalar SL
field, is defined \cite{MundRS17a} as
$a(x,e) := - m^{-1}\,\del^\mu A_\mu(x,e)$. For bosons of higher spin,
one can define a family of escorts recursively, as 
follows~\cite{MundRS17b}. For a spin~$s$ field, one first declares the
string-localized potential
$a^{(s)}_\muu(x,e) \equiv A_\muu(x,e) := I_e^s F_{[\muu,\nuu]}(x)$ as
in~\eqref{eq:high-potential}. For $r = s-1,\dots,1,0$, one defines
escort fields $a^{(r)}$ by putting
$$
a^{(r)}_{\mu_1\cdots\mu_r}(x,e) 
:= - m^{-1}\, \del^\nu a^{(r+1)}_{\mu_1\cdots\mu_r\nu}(x,e).
$$
Thus a spin-$2$ potential $A_{\mu\nu}(x,e)$ has two escorts, a vector
$a^{(1)}_\mu(x,e)$ and a scalar $a^{(0)}(x,e)$.

These escort fields obey the axial condition and the field equations:
\begin{equation}
e^\mu a^{(r)}_{\mu\kaa} = 0, \quad
\del^\mu a^{(r)}_{\mu\kaa} = - m a^{(r-1)}_\kaa, \quad
g^{\mu\nu} a^{(r+1)}_{\mu\nu\kaa} = - a^{(r-1)}_\kaa,
\label{eq:walk-with-me} 
\end{equation}
where $\kaa = \ka_2\cdots\ka_r$ denotes the other indices. One can 
recover a point-like potential from $A_\muu(x,e)$ and derivatives of 
its escorts. For instance, for spin~$2$,
$$
A^\pt_{\mu\nu}(x) := A_{\mu\nu}(x,e) - \frac{1}{m}
\bigl( \del_\mu a^{(1)}_\nu(x,e) + \del_\nu a^{(1)}_\mu(x,e) \bigr)
+ \frac{1}{m^2}\, \del_\mu \del_\nu a^{(0)}(x,e).
$$

Now, $A_\muu(x,e)$ and its escorts have well-defined massless limits
as $m \downto 0$, but do not yield a decoupling of the lower
helicities $h = \pm r$ without further modification. Rewrite 
\eqref{eq:potential-player} as an operator in $x$-space:
$$
E_{\mu\nu}(e,e) 
:= g_{\mu\nu} + (e_\nu\,\del_\mu + e_\mu\,\del_\nu) I_e
+ e^2 \del_\mu \del_\nu I_e^2 \,.
$$
On replacing $A_{\mu\nu}(x,e)$ by
$$
A^{(2)}_{\mu\nu}(x,e) 
:= A_{\mu\nu}(x,e) + \half E_{\mu\nu}(e,e)\, a^{(0)}(x,e),
$$
one gets a potential that in the massless limit is traceless,
decouples from $a^{(1)}_\mu$ and~$a^{(0)}$, and has the correct
two-point function $M_0^{A^{(2)}A^{(2)}}$. For higher spins,
analogously modified potentials yield the desired decoupling of lower
helicities~\cite{MundRS17b}.

\section{The principle of string independence: flavourdynamics}
\label{sec:Epaminondas}

Not so long ago one could read in the short version of the Review of
Particle Physics \cite[p.~209]{RPPhlittle16} that the fermions of
the~SM are chiral, and that therefore in principle they should be
massless. Furthermore, given that they are in fact massive, somehow a
``spontaneous symmetry breaking'' generating their masses must have
happened. Fortunately, these disquisitions appear to have vanished in
more recent booklets~\cite{RPPhlittle20}. For that is to have matters
backwards. To begin with, what may or may not be chiral are the
\textit{interactions}; quarks are not schizophrenic, chiral in
flavourdynamics and non-chiral in chromodynamics: chirality is not a
Platonic property of particles. Secondly, our work together with
Mund~\cite{Rosalind} established that the couplings of the massive
vector bosons with fermions are chiral precisely because the former
are massive. Our finding is a \textbf{theorem}, rigorously derived
from the string independence principle at second order in $\bS$-matrix
theory: given merely the masses and charges of the interaction
carriers and the fermions, \textit{all of electroweak dynamics flows
from this principle}. For lack of space, here we just provide an
outline of the construction, leaving the full derivation of the model
for~\cite{Barbara}.

\subsection{The intermediate bosons}
\label{ssc:paria-delicta}

Formula \eqref{eq:Faraday} for a set of four Faraday fields
$F_a = \set{W^\pm,Z,\ga}$ is our starting point. It proves useful to
consider the spinless string-local \textit{escort} fields:
\begin{equation}
\phi_b(x,e) := \sum_r \int d\mu(p)\, \biggl[
e^{i(px)} \frac{i(\eps_r(p)\,e)}{(pe) + i\eps}\, a_{r,b}^\7(p)
+ e^{-i(px)} \frac{-i(\eps_r(p)\,e)^*}{(pe) - i\eps}\, a_{r,b}(p) \biggr].
\label{eq:hire-an-escort} 
\end{equation}
As remarked already, the differences
$$
A^\mu_b(x,e) - \del^\mu \phi_b(x,e) =: A^{\pt,\mu}_b(x)
$$
define pointlike \textit{Proca} fields, so that $dA^\pt_b = F_b$. All
these fields live on the same Fock spaces as the $F_b$ and have the
same mass. Moreover:
$$
\phi_b(x,e) = \int_0^\infty A^{\pt,\la}_b(x + se) e_\la \,ds.
$$
Note the relations $(e\,\del\phi_b) = -(eA^\pt_b)$ and
$\del_\mu A_b^\mu(x,e) + m_b^2 \phi_b(x,e) = 0$, in accordance
with~\eqref{eq:walk-with-me}. The second one follows directly
from~\eqref{eq:Faraday} and~\eqref{eq:hire-an-escort}, since
$(p\,\eps_r(p)) = 0$.

Now, a complete account of electroweak theory would start by showing
that, when the string independence principle is applied to the
physically relevant set of boson SLFs, with their known masses and
charges, replacing the standard pointlike fields, plus one
\textit{physical} Higgs%
\footnote{Following Okun and for obvious grammatical reasons,
henceforth we refer to a (physical) Higgs boson as a higgs, with a
lower-case~h.}
particle $\phi_4(x)$, one recovers precisely the phenomenological
couplings of flavourdynamics in the SM, with massive bosons mediating
the weak interactions, and the $U(2)$ structure constants, as
expounded for instance in~\cite{Scheck12}. For want of space, we just
summarize here the main conclusions concerning the boson sector.

\begin{itemize}

\item
Apart from the higgs particle sector, a string-local theory of
interacting bosons at first order in the coupling constant~$g$ -- see
Eq.~\eqref{eq:Dyson-sphere} below -- must be of the form:
\begin{align}
S^B_1(x,e) &= g \sump_{a,b,c} f_{abc} (m^2_a - m_b^2 - m_c^2)
\bigl( A_a(x,e) A_b(x,e) \phi_c(x,e)
\notag \\
&\quad - A_a(x,e)\,\del\phi_b(x,e) \phi_c(x,e) \bigr)
+ g \sum_{a,b,c} f_{abc} F_a(x) A_b(x,e) A_c(x,e),
\label{eq:bosons-mate} 
\end{align}
where we omit the notation $\wick:\text{\,---\,}:$ for Wick products,
and the restricted sum $\sum'$ runs over massive fields only. Here the
$f_{abc}$ denote the (completely skewsymmetric) structure constants of
the (reductive) symmetry group of the model; the mass of the vector
boson $A_a$ is denoted $m_a$, and complete contraction of Lorentz
indices is understood. Notice that the escort fields hold a place
somewhat analogous to St\"uckelberg fields.

\item
It is straightforward to check that the $1$-form $d_l S^B_1$,
measuring the dependence on the string variable of the vertices
in~\eqref{eq:bosons-mate}, is a divergence: 
$d_l S^B_1(x,l) = \bigl(\del Q^B_1 \bigr)(x,l)$, where $Q^B_{1\mu}$
is given by:
\begin{equation}
2g \sum_{a,b,c} f_{abc} (F_a A_c)_\mu w_b 
+ g \sump_{a,b,c} f_{abc} (m_a^2 + m_c^2 - m_b^2)
(A_{a\mu} - \del_\mu\phi_a) \phi_c w_b.
\label{eq:bosons-shadow} 
\end{equation}
We did introduce above the form-valued fields $w_a := d_e\phi_a$. We
need $Q^B_1$ to prove chirality of the couplings to the fermion
sector.

\item
At once we adapt our notation to the one used in the SM. This model
has three nonzero masses $m_1 = m_2 < m_3$ and one $m_4 = 0$. Defining
the Weinberg angle%
\footnote{This makes sense in the normalized theory
\cite[Chap.~29]{Schwartz14}.}
$\Th$ by $m_1/m_3 =: \cos\Th$, we employ the basis in which
$$
f_{123} = \half \cos\Th, \quad  
f_{124} = \half \sin\Th, \quad  f_{134} = f_{234} = 0,
$$
all other $f_{abc}$ following from complete skewsymmetry. They are
seen to be the structure constants of (the Lie algebra of) the $U(2)$
determined by the \textit{physical} particle fields. We use the
standard notations
$$
W_\pm \equiv \frac{1}{\sqrt{2}}(W_1 \mp iW_2)
:= \frac{1}{\sqrt{2}}(A_1 \mp iA_2), \quad
Z := A_3,  \quad  A := A_4  
$$
and similarly for $\phi_\pm$, $w_\pm$, $\phi_Z$ and~$w_Z$; with masses
$m_W = m_1$, $m_Z = m_3$ and $m_\ga = m_4 = 0$.

\item
With this in hand, we focus on \eqref{eq:bosons-shadow}, keeping in
mind that, although the photon has no escort field, the field $w_4$
exists at the same title as $w_1$, $w_2$, $w_Z$. The first summand
in~\eqref{eq:bosons-shadow} yields:
\begin{align*}
& 2g \tsum f_{abc}(\del_\mu A_{a\la} - \del_\la A_{a\mu}) A_c^\la w_b
\\
&\quad = ig \sin\Th \bigl[ (\del_\mu A_\la - \del_\la A_\mu)
(w_- W_+^\la - w_+ W_-^\la)
+ (\del_\mu W_{-\la} - \del_\la W_{-\mu})(w_+ A^\la - w_4 W_+^\la)
\\
&\hspace*{6em} 
+ (\del_\mu W_{+\la} - \del_\la W_{+\mu})(w_4 W_-^\la - w_- A^\la)
\bigr] 
\\
&\quad + ig \cos\Th \bigl[(\del_\mu Z_\la - \del_\la Z_\mu)
(w_- W_+^\la - w_+ W_-^\la) 
+ (\del_\mu W_{-\la} - \del_\la W_{-\mu})(w_+ Z^\la - w_Z W_+^\la) 
\\
&\hspace*{6em}
+ (\del_\mu W_{+\la} - \del_\la W_{+\mu})(w_Z W_-^\la - w_- Z^\la)
\bigr].
\end{align*}

\item
Our $Q^B_{1\mu}$ above is incomplete, since bosonic couplings
involving the higgs sector are not included. They are also derived
from the string independence principle. Of those, for our purposes in
this analysis we need only
$$
\frac{g}{2\cos\Th} m_Z (\phi_4 (\del_\mu\phi_Z - Z_\mu)
- \del_\mu\phi_4\, \phi_Z) w_Z;
$$
actually these play a pivotal role in our work in~\cite{Rosalind}.
Clearly, terms of this type are suggested by the last group of
summands in \eqref{eq:bosons-shadow}.

\item
The expected $g^2 AAAA$ terms and thus the trappings of the
geometrical gauge approach are recovered in our formalism from string
independence at the level of~$S_2$.
\end{itemize}

\subsection{On string independence}
\label{ssc:in-maleficiis}

The general procedure to be followed engages the construction of the
Bogoliubov-type functional scattering operator $\bS[g;c]$ dependent on
a multiplet~$g$ of external fields and a test function $c \in \sD(H)$
with integral equal to~$1$ that averages over the string
directions~\cite{Gass22}. Bogoliubov's $\bS[g;c]$ obeys the customary
conditions of causality, unitarity and covariance. One looks for it as
a formal power series in~$g$,
\begin{equation}
\bS[g;c] := 1 + \sum_{n=1}^\infty \frac{i^n}{n!} \prod_{k=1}^n
\prod_{l=1}^m \int d^4x_k \int d^4(e_{k,l})\,g(x_k)\,c(e_{k,l})\,
S_n(x_1,\ee_1;\dots;x_n,\ee_n).
\label{eq:Dyson-sphere} 
\end{equation}
Only the first-order vertex coupling $S_1 = S_1(x,\ee)$, a Wick
polynomial in the free fields, is postulated -- already under severe
restrictions. It depends on an array $\ee = (e_1,\dots,e_m)$ of string
coordinates, with $m$ the maximum number of SLFs appearing in a
sub-monomial of~$S_1$.

For $n \geq 2$, the $S_n$ are time-ordered products that must be
recursively constructed. These $S_n$ will be symmetric with respect to
permuting the string coordinates, which are smeared with the
\textit{same} test function~$c$. (This symmetry will be heavily
exploited in what follows.) The extension of the $S_n$-products across
exceptional sets like
$$
\bD_2 := \set{(x,\ee;x',\ee') : (x + \bR^+ e_k) \cap (x' + \bR^+ e'_l)
\neq \emptyset \text{ for some } k,l},
$$
and similar ones~$\bD_n$, is the normalization problem in a nutshell.

The hypothesis of perturbatively interacting SLF theory is simple
enough: physical observables and closely related quantities,
particularly the $\bS$-matrix, cannot depend on the string
coordinates. This is the -- intrinsically quantum -- \textbf{string
independence} principle, which here replaces the gauge ``principle''
and classical Lagrangians with advantage.

\subsection{Summary of the proof of chirality}
\label{ssc:mutua-pensatione}

The couplings between interaction carriers and matter currents in a
theory with massive or massless vector bosons $A_{a\mu}$ must be of
the form
\begin{gather}
g\bigl (b^a A_{a\mu} J_V^\mu + \tilde b^a A_{a\mu} J_A^\mu 
 + c^a \phi_a S + \tilde c^a \phi_a S_5 \bigr) ; 
\label{eq:tapa-del-perol} 
\\
\shortintertext{where}
J_V^\mu = \ovl\psi \ga^\mu \psi, \quad
J_A^\mu = \ovl\psi \ga^\mu \ga^5 \psi, \quad 
S = \ovl\psi \psi, \quad 
S_5 = \ovl\psi \ga^5 \psi,
\notag
\end{gather}
with electric charge conserved in the interaction vertices. Our key
assumption is that these $A_a^\mu$ and~$\phi_a$ above are now given as
string-local quantum fields, thus satisfying renormalizability by
power counting. There are \textit{no other scalar couplings which
comply with renormalizability}. To wit, Lorentz invariance requires
that all cubic terms be of the above form, and renormalizability
forbids quartic terms.%
\footnote{Since the two Fermi fields required by Lorentz invariance
already have scaling dimension~$3$, any two further fields would
give~$5$, exceeding the power-counting limit.}

The $\psi$ in \eqref{eq:tapa-del-perol} are ordinary fermion fields --
we should not assume chiral fermions \textit{ab~initio}, and we
do~not.

The coefficients $b^a$, $\tilde b^a$, $c^a$, $\tilde c^a$
in~\eqref{eq:tapa-del-perol} are to be determined from string
independence.

\medskip

With the above in the bag, one just makes the most general Ansatz of
the kind \eqref{eq:tapa-del-perol}. At first order, the string
independence requirement amounts to hermiticity, which demands for the
couplings with fermions:
\begin{align*}
S_1^F(x,l) &= g\bigl( b_1 W_{-\mu} \bar e \ga^\mu \nu 
+ \tilde b_1 W_{-\mu} \bar e \ga^\mu \ga^5 \nu
+ b_1 W_{+\mu} \bar\nu \ga^\mu e
+ \tilde b_1 W_{+\mu} \bar\nu \ga^\mu \ga^5 e
\\
&\qquad + b_3 Z_\mu \bar e \ga^\mu e
+ \tilde b_3 Z_\mu \bar e \ga^\mu \ga^5 e
+ b_4 Z_\mu \bar\nu \ga^\mu \nu
+ \tilde b_4 Z_\mu \bar\nu \ga^\mu \ga^5 \nu
+ b_5 A_\mu \bar e \ga^\mu e
\\
&\qquad + i(m_e - m_\nu) b_1 \phi_- \bar e \nu 
+ i(m_e + m_\nu) \tilde b_1 \phi_- \bar e \ga^5 \nu 
- i(m_e - m_\nu) b_1 \phi_+ \bar\nu e 
\\
&\qquad + i(m_e + m_\nu)\tilde b_1 \phi_+ \bar\nu \ga^5 e
+ 2im_e \tilde b_3 \phi_Z \bar e \ga^5 e 
+ 2im_\nu \tilde b_4 \phi_Z \bar\nu \ga^5 \nu
\\
&\qquad + c_0 \phi_4 \bar e e + \tilde c_0 \phi_4 \bar e \ga^5 e 
+ c_5 \phi_4 \bar\nu \nu + \tilde c_5 \phi_4 \bar\nu \ga^5 \nu \bigr),
\end{align*}
where $\phi_\pm,\phi_Z$ are escort fields and $\phi_4$ is the higgs
field. Here $e$ stands for an electron or muon or $\tau$-lepton
pointlike field or for a suitable combination of the quarks $d,s,b$;
and $\nu$ for the neutrinos or for a suitable combination of the
quarks $u,c,t$ -- it is enough to consider just one generation of
leptons.

After harder work, second-order string independence allows for the
full determination of the fermionic couplings:
\begin{align*}
S_1^F = g \biggl\{ 
& - \frac{1}{2\sqrt{2}}\, W_{-\mu} \bar e \ga^\mu (1 - \ga^5) \nu
- \frac{1}{2\sqrt{2}}\, W_{+\mu} \bar\nu \ga^\mu (1 - \ga^5) e
+ \frac{1 - 4\sin^2\Th}{4\cos\Th}\, Z_\mu \bar e \ga^\mu e
\\
& - \frac{1}{4\cos\Th}\, Z_\mu \bar e \ga^\mu \ga^5 e
- \frac{1}{4\cos\Th}\, Z_\mu \bar\nu \ga^\mu (1 - \ga^5) \nu
+ \sin\Th\, A_\mu \bar e \ga^\mu e
\\
& + i \frac{m_e - m_\nu}{2\sqrt{2}}\,
(\phi_- \bar e \nu - \phi_+ \bar\nu e)
- i \frac{m_e + m_\nu}{2\sqrt{2}}\, 
(\phi_- \bar e \ga^5 \nu + \phi_+ \bar\nu \ga^5 e)
\\
& - i \frac{m_e}{2\cos\Th}\, \phi_Z \bar e \ga^5 e 
+ i \frac{m_\nu}{2\cos\Th}\, \phi_Z \bar\nu \ga^5 \nu
+ \frac{m_e}{2m_W}\, \phi_4 \bar e e
+ \frac{m_\nu}{2m_W} \phi_4 \bar\nu \nu \biggr\}.
\end{align*}
(As established in~\cite{Rosalind}, too, one can sweep away the escort
fields. However, this is not convenient in the present dispensation.)

\section{The principle of string independence: gluodynamics}
\label{sec:contentus}

Let us now reconsider QCD from our present viewpoint. Suppose that we
are given three massless fields $A_{\mu a}$ ($a = 1,2,3$). Essentially
following \cite{Borisov}, we first show that, for a mutual cubic
coupling modulo divergences, string independence \textit{at first
order} in the coupling constant enables \textbf{only} the Wick-product
combination:
\begin{equation}
S_1(x,e_1,e_2) = \frac{g}{2}\,
f_{abc}\, A_{\mu a}(x,e_1) A_{\nu b}(x,e_2) F^{\mu\nu}_c(x),
\label{eq:first-cousin} 
\end{equation}
where the $f_{abc}$ are \emph{completely skewsymmetric} coefficients.
(Subindices that appear twice are summed over, and normal ordering is
understood.) For the physics of the model described by $S_1$ in
Eq.~\eqref{eq:first-cousin} to be string-independent, one must require
that a vector field $Q^\mu(x,\ee)$ exist which, after symmetrization
in the string variables, fulfils
$$
d_{e_1} S_1^\sym(x,e_1,e_2) = (\del Q) := \del_\mu Q^\mu,
$$
so that on applying integration by parts in the ``adiabatic limit'' as
$g$ goes to a set of constants, the contribution from the divergence
vanishes. Similarly at higher orders, as we shall see. Then, as the
covariant functional $\bS[g;c]$ approaches the invariant physical
scattering matrix~$\bS$, so $U(a,\La)\, \bS\, U^\7(a,\La) = \bS$, all
dependence on the strings disappears.

We call ``gluons'' the fields appearing in Eq.~\eqref{eq:first-cousin}
and shall show from string independence that they behave precisely 
as~such.

\subsection{Complete skewsymmetry of the $f_{abc}$}
\label{ssc:accusare}

Before proceeding, we wish to mention an important paper
\cite{AsteS99} by Aste and Scharf, wherein the reductive Lie algebra
structure of Yang--Mills theory was derived from a \textit{quantum}
field formulation for gauge invariance. That paper is representative
of the work by the Zurich school which, on the basis of BEG theory,
too, tried to liberate quantum field theory from both the tyranny of
classical concepts and the perplexities of gauge invariance, with the
help of cohomological methods -- see also in this regard
\cite{DuetschS99} and \cite{Ausonia}, the latter by one of us. Here we
go further, in that perturbatively keeping positivity, locality and
covariance of the interactions in our purely quantum, bottom-up
approach allows one to recover all the features of gluodynamics.

That our claims can be sustained was already proved in relation to
perturbative QCD in~\cite{Borisov}. Thanks to more recent
work~\cite{Gass22}, here we are able to sketch improvements in the
technical details. Note first from the simplest version of
Eq.~\eqref{eq:great-potential} that the string differential of the
gluon field is a gradient:
\begin{align}
d_e A_{\mu b}(x,e) := i\del_\mu \sum_{h=\pm} 
& \int \! d\mu(p)\, d\eps^\rho\, \biggl[ e^{i(px)} \biggl(
\frac{\eps_\rho^h}{(pe)} - \frac{p_\rho (\eps^h e)}{(pe)^2} \biggr)
a^\7_b(p,h) 
\nonumber \\
& - e^{-i(px)} \biggl( \frac{\eps_\rho^h}{(pe)}
- \frac{p_\rho(\eps^h e)}{(pe)^2} \biggr)^* a_b(p,h) \biggr]
=: \del_\mu u_b(x,e).
\label{eq:gluehwein} 
\end{align}
Let us examine the general cubic Ansatz renormalizable by power
counting. With an obvious notation:
\begin{align*}
S'_1(x,e_1,e_2,e_3) 
&:= g f^1_{abc}\, A_{\mu a}(x,e_1) A_{\nu b}(x,e_2)\, 
\del^\mu A^\nu_c(x,e_3) 
\\
& =: gf^1_{abc}\,A_{\mu a}^1 A_{\nu b}^2 \del^\mu A^3_{\nu c} 
:= g(d_{abc} + l_{abc})\, 
A_{\mu a}^1 A_{\nu b}^2 \del^\mu A^{3\nu}_c,
\end{align*}
where $d_{abc} = d_{acb}$ and $l_{abc} = -l_{acb}$. Because
$\del^\mu A^1_{\mu a} = 0$, on symmetrizing $e_2 \otto e_3$ we can
drop a total divergence and are left with
\begin{align*}
S''_1(x,e_1,e_2,e_3) 
:= g f^2_{abc}\, A_{\mu a}(x,e_1) A_{\nu b}(x,e_2)\,
\del^\mu A^\nu_c(x,e_3),
\end{align*}
where $f^2_{abc}$ is skewsymmetric under $b \otto c$. Repeating the
procedure, we split $f^2_{abc}$ into a part $f^+_{abc} = f^+_{bac}$
and a totally skewsymmetric part $f^-_{abc}$, and study
$$
S''^-_1(x,e_1,e_2) := \frac{g}{2} f^-_{abc} 
A_{\mu a}(x,e_1) A_{\nu b}(x,e_2) F_c^{\mu\nu}(x).
$$
Now, in view of~\eqref{eq:gluehwein}:
\begin{align*}
d_{e_1} {S''_1}^-(x,e_1,e_2) 
= \frac{g}{2}\, f^-_{abc}\,\del_\mu(u^1_a A^2_{\nu b} F_c^{\mu\nu});
\end{align*}
so this part \textit{is} string-independent in the adiabatic limit. On
the other hand, it is easily seen that the $f^+_{abc}$ must vanish
identically for the symmetric part to be string-independent. In
conclusion, string independence at first order demands
\textit{complete skewsymmetry} of the~$f_{abc}$.

\subsection{String-independence at second order}
\label{sec:aegri}

As yet the $f_{abc}$ do not make a Lie algebra. For that a
\textit{Jacobi identity} is required. We obtain it from string
independence of the functional $\bS$-matrix at second order in the
couplings. Since time-ordering products need to be defined here, that
requires a normalization procedure.

Let $y_i = x_i + s_i e_i$ and $y'_j = x'_j + s'_j e'_j$. The
second-order \textit{tree} graph contribution to the $\bS$-matrix is
given by the string integral:
$$
\T[S_1(x,e_1,e_2)\, S_1(x',e'_1,e'_2)]\bigr|_\tree
= \int_0^\infty ds_1\cdots ds'_2\,
\T[S_1(x,y_1,y_2)\, S_1(x',y'_1,y'_2)]\bigr|_\tree \,.
$$
Since the gluon propagators are diagonal in the color indices, this 
simply takes the form
$$
\T[S_1(x,e_1,e_2)\, S_1(x',e'_1,e'_2)]\bigr|_\tree
= \sum_{\vf,\chi'} \frac{\del S_1(x_1,y_1,y_2)}{\del\vf}\,
\vev{\T \vf\,\chi'} \,\frac{\del S'_1(x_1,y'_1,y'_2)}{\del\chi'}
$$
for the fields $\vf,\chi$ entering $S_1$; Wick ordering is understood
throughout. Inserting the explicit formulas for our problem, we 
obtain:
\begin{align}
\quarter g(x) g(x') f_{abc} f_{a'b'c'}
& \Bigl[ \vev{\T F^{\mu\nu}_c F'^{\ka\la}_{c'}} A^1_{\mu a}
A^2_{\nu b}{A'}^{1'}_{\ka a'}{A'}^{2'}_{\la b'}
\label{eq:to-be-symmetrized} 
\\
&+ \vev{\T A^1_{\mu a} F'^{\ka\la}_{c'}} A^2_{\nu b}
F^{\mu\nu}_c{A'}^{1'}_{\ka a'}{A'}^{2'}_{\la b'} + (e_1 \otto e_2)
\notag
\\
&+ \vev{\T F^{\mu\nu}_c{A'}^{1'}_{\ka a'}}A^1_{\mu a} A^2_{\nu b}
A^2_{\la b'}F'^{\ka\la}_{c'} + (e'_1\leftrightarrow e'_2)
\notag
\\ 
&+ \bigl( \vev{\T A^1_{\mu a}A'^{1'}_{\ka a'}} A^2_{\nu b}
F^{\mu\nu}_c {A'}^{2'}_{\la b'} F'^{\ka\la}_{c'}
+ (e_1 \otto e_2)\bigr) + (e'_1 \otto e'_2) \Bigr].
\notag
\end{align}

The propagators in the previous formula are unspecified as yet. For
the purpose one naturally considers in the first place the
``kinematic'' propagators:
\begin{equation}
\vev{\T_0 \vf(x,e)\,\chi(x',e')} := \frac{i}{(2\pi)^4} \int d^4p\,
\frac{e^{-i(p(x - x'))}}{p^2 + i0}\, M^{\vf\chi}(p,e,e'),
\label{eq:pandemonium} 
\end{equation}
where the $M^{\vf\chi}(p,e,e')$ are given by $M^{\vf\chi}_{*\8} 
:= \sum_\sg \ovl{u^{\sg;\vf}_*(p,e)}\, u^{\sg;\chi}_\8(p,e')$ for the
appropriate spacetime indices $*,\8$ on the respective intertwiners.
Besides \eqref{eq:field-of-play}, one calls for:
$$
M^{FA}_{\mu\nu,\la}(p,e')
= i\biggl( p_\mu g_{\nu\la} - p_\nu g_{\mu\la} 
+ p_\la \frac{p_\nu e'_\mu - p_\mu e'_\nu}{(pe') + i0} \biggr),
$$
to be found for instance in our~\cite{Rosalind}. The more singular
$M^{AA}$ is considered in Appendix~\ref{app:satis}. Note that in all
generality
$$
d_e \vev{\T_0 \vf(x,e)\, \chi(x',e')}
= \vev{\T_0 d_e\vf(x,e)\, \chi(x',e')},
$$
and similarly for~$d_{e'}$. In general one should admit for the
propagators:
\begin{align*}
\vev{\T F_c^{\mu\nu}{F'}_d^{\ka\la}} 
&= \dl_{cd} \bigl[ \vev{\T_0 F^{\mu\nu} {F'}^{\ka\la}}
+ c(\eta^{\mu\ka} \eta^{\nu\la}
- \eta^{\mu\la} \eta^{\nu\ka}) \,\dl(x - x') \bigr];
\\
\vev{\T F_c^{\mu\nu} {A'}^{1'\ka}_d} 
&= \dl_{cd} \bigl[ \vev{\T_0F_c^{\mu\nu}{A'}^{1'\ka}}
+ c(\eta^{\mu\ka} e'^\nu_1 
- \eta^{\nu\ka} e'^\mu_1) I_{-e'_1} \dl(x - x') \bigr]; 
\\
\vev{\T A_a^{1\mu} F'^{\ka\la}_d}
&= \dl_{ad} \bigl[ \vev{\T_0 A_a^{1\mu} F'^{\ka\la}_d}
+ c(\eta^{\mu\ka} e^\la_1 - \eta^{\mu\la} e^\ka_1)
I_{e_1} \dl(x - x') \bigr];
\\
\vev{\T A_a^{1\mu} A_d'^{1'\ka}}
&= \dl_{ad} \bigl[ \vev{\T_0 A_a^{1\mu} A_d'^{1'\ka}}
+ c(\eta^{\mu\ka}(e_1e'_1) - e'^\mu_1 e^\ka_1)
I_{e_1} I_{-e'_1} \dl(x - x') \bigl],
\end{align*}
for an indeterminate constant~$c$. However, we do not need the 
non-kinematic terms, which are killed by the string independence 
principle \cite[Sect.~6.1.2]{GassDiss22}.

Looking back at Eq.~\eqref{eq:to-be-symmetrized}, we summarize what we
have obtained so far:
\begin{align}
\MoveEqLeft{%
\T_0[S_1(x,e_1,e_2) S_1(x',e'_1,e'_2)] \bigr|_{\sym,\tree}
= \quarter g(x) g(x') f_{abc} f_{def}}
\notag \\
& \x \Bigl[ \vev{T_0 F_c^{\mu\nu} F'^{\mu\nu}_f} 
A^1_{\mu a} A^2_{\nu b} A'^{1'}_{\ka d} A'^{2'}_{\la e}
+ 2\vev{T_0 A^1_{\mu a} F'^{\ka\la}_f} A^2_{\nu b} F^{\mu\nu}_c
A'^{1'}_{\ka d} A'^{2'}_{\la e}
\notag \\
&\quad + 2 \vev{T_0 F^{\mu\nu}_c A'^{1'}_{\ka d}} A^1_{\mu a}
A^2_{\nu b} A'^{2'}_{\la e} F'^{\ka\la}_f
+ 4 \vev{T_0 A^1_{\mu a} A'^{1'}_{\ka d}} A^2_{\nu b} F^{\mu\nu}_c
A'^{2'}_{\la e} F'^{\ka\la}_f \Bigr]_\sym.
\label{eq:deficit} 
\end{align}

Now Eq.~\eqref{eq:gluehwein} entails
$d_{e_1}\vev{T_0\,A^1_{\mu a} -} = \del_\mu\vev{T_0\,u^1_a -}$.
As a consequence, most of the terms contained in \eqref{eq:deficit}
are total divergences and can be dropped. In the
end~\cite{GassDiss22}, one is left with:
\begin{align*}
& d_{e_1} \T_0[S_1(x,e_1,e_2)\, S_1(x',e'_1,e'_2] \bigr|_{\sym,\tree}
= - \frac{g^2(x)}2 f_{abc} f_{dec} d_{e_1}
\bigl[ (A^1_a A^{2'}_d) (A^2_b A^{1'}_e)\bigr]_\sym
\\
&- \frac{g^2(x)}{6} \bigl[ f_{abc} f_{dec} + f_{adc} f_{ebc}
+ f_{aec} f_{bdc} \bigr]
\bigl( u^1_a F^{\ka\la}_b A^{1'}_{\ka d} A^{2'}_{\la e}
+ \text{two similar terms} \bigr).
\end{align*}
The first one yields the quartic term in Yang--Mills theory,
redefining~$S_2$. The second obstruction needs to vanish in order to
achieve string independence, yielding the Jacobi identity.

\medskip

A contemporaneous, even more elegant bottom-up approach without prior
prejudices neither about Lagrangians nor classical gauge fields, also
leading uniquely to the Yang--Mills type theory, is found in
\cite[Chap.~25]{Schwartz14}. This one utilizes the spinor-helicity
formalism. Far from Yang's Platonic dictum ``symmetry dictates
interaction'', we do concur with that approach, in an Aristotelian
mould, that \textit{interaction dictates symmetry}.

\medskip

Compared with this latter approach, as well as the standard one, the
main drawback of string-local field theory is manifest in formulas
like \eqref{eq:potential-player}: calculations quickly become rather
complicated by the abundance of propagator terms. At present,
tree-graph calculations have been pushed reasonably far
\cite{Tippner19}; and only a few loop-graph calculations with stringy
particles on internal lines have been tamed.

\section{On the scalar particle(s)}
\label{sec:princeps}

The attentive reader of Section~\ref{sec:Epaminondas} will have
learned the important role that the scalar ``escort fields''
associated to the \textit{massive} vector bosons of electroweak theory
play, as a bridge between the SLF vector fields representing them and
the corresponding Proca fields. The higgs field there did
inauspiciously make its appearance as a kind of escort field
associated to the fourth carrier of the electroweak theory, the
photon. There string independence quickly establishes the need for at
least one scalar particle whose coupling to the particles of the SM is
proportional to their masses. As remarked first by Alejandro Ibarra
\cite[Sect.~7.2]{Rosalind}, its presence warrants chirality of the
interactions of the charged bosons right away.

One also reads in \cite[p.~210]{RPPhlittle16}: ``The higgs boson
couplings are not dictated by any local gauge symmetry''. We already
showed the derivative character of the latter concept. String-local
theory, on the other hand, recognizes that the higgs's interactions
are on the same footing as those of any other particles. One finds,
too, in a Review of Particle Physics \cite[p.182]{RPPh18}: ``\dots\
the SM Higgs couplings to fundamental fermions are linearly
proportional to the fermion masses, whereas the couplings to bosons
are proportional to the square of the boson masses''.
Unfortunately, this is reproduced verbatim in the newer
\cite[p.~205]{RPPh20}. Now, if one naturally uses the overall
electroweak coupling constant, instead of the ``Fermi constant''
associated to the intangible ``spontaneous symmetry breaking
mechanism'', mere dimension counting tells us that the couplings of
scalar particle(s) to \textit{any others} must be proportional to
their masses. From the viewpoint of modular localization and Wigner
particle theory and from plain rationality, the above statement in
\cite{RPPh18, RPPh20} makes little sense.

Now, we ought to able to recover \textit{all} the higgs' couplings,
particularly the self-couplings, from string independence. As it turns
out, this requires examination of third-degree couplings. Our task is
helped by having done so already in the past \cite{Ausonia, Aurora},
in the framework of causal gauge invariance. The method and the
results, showing again the efficient detective character of the string
independence principle, will appear in the review paper~\cite{Barbara}.%
\footnote{Should there be more scalars in nature~\cite{BieotterHW22},
one still learns much about their couplings from SLF theory. But it
would no longer be possible to determine those couplings entirely.}

\subsection*{Acknowledgements}

Both J.M.G-B. and J.C.V. were supported by the research project
``Nuevos aspectos de la física cuántica'', \#C1--051 of the
Universidad de Costa Rica. J.M.G-B. was also supported by the research
project ``Quantum gravity phenomenology in the multi-messenger
approach'', European Cooperation COST Action CA18108. We heartily
thank the anonymous referee, whose comments helped to improve the
paper.

\appendix

\section{Microlocal analysis of string-localized theories}
\label{app:satis}

In subsection \ref{ssc:in-maleficiis} the first-order vertex coupling
$S_1(x,\ee) = S_1(x,e_1,\dots,e_m)$ that enters the Dyson expansion
\eqref{eq:Dyson-sphere} depends on a spacetime point~$x$ and several
strings $e_l$, one for each string-like field in that Wick polynomial;
see~\eqref{eq:first-cousin}, for instance. One should not try to
simplify by taking all the~$e_l$ to be equal, since this may lead to
ill-defined products of distributional terms such as
$\vev{\T\vf\,\chi'}$.

A careful analysis of the singularities affecting the extension of 
time-ordered products needed to construct $S_2$ and higher-order terms
has been carried out recently by Ga{\ss} in~\cite{Gass22}. We 
briefly review what is involved.

Two distributions $u$ and~$v$, defined on some open set
$X \subseteq \bR^n$, can be multiplied if their wave-front sets do not
clash. The wave-front set $WF(u)$ is the set of covectors $(x,p)$ for
which $u$ is not smooth at $x \in X$ and $p \neq 0$ is a direction
along which the Fourier transform $\hat u$ does not have rapid decay.
If there are $(x,p) \in WF(u)$ and $(x,k) \in WF(v)$ such that
$k = -p$, then the distributional product $uv$ is absent; otherwise,
$uv$ is well defined and $WF(uv)$ is $W(u) \cup W(v)$ together with
all such $(x, p + k)$. For the full story of WF sets and their
properties, see~\cite{Hormander90}.

An easy pair of examples of wave-front sets, for $X = \bR$, is
\begin{equation}
WF((t \pm i0)^{-1}) = \set{(0,\la) : \la \gtrless 0},
\label{eq:one-diml-wave-fronts} 
\end{equation}
since $(t \pm i0)^{-1}$ have the Fourier transforms $\theta(\pm\la)$.
From there, WF sets of other distributions, such as
$(p^2 \pm i0)^{-1}$ or $(p^2 - m^2 \pm i0)^{-1}$ in $p$-space, can be
found by pullback operations.

To deal with string-local fields, we put
$u_\pm(p,e) := [(pe) \pm i0]^{-1}$ for $(p,e) \in M^4 \x H$, already
considered, whose wave-front sets are found to be~\cite{Gass22}:
$$
WF(u_\pm) = \set{(p,e; \la e, \la p)
: (pe) = 0, \ e^2 < 0, \ \la \lessgtr 0}.
$$
These signs of~$\la$ are opposite to those
of~\eqref{eq:one-diml-wave-fronts} because we use the mostly-negative 
metric on~$M^4$. Because of the fixed sign of~$\la$, it is seen that 
there exist tempered distributions $(u_+)^k$ and $(u_-)^k$ for any
$k = 2,3,\dots$ but that the product $u_+\,u_-$ does \textit{not} 
exist.

In particular, when computing propagators for string-local potentials
$A_\mu(x,e)$, the expression \eqref{eq:potential-player} contains a
term $(ee') p_\mu p_\rho\, u_-(p,e)\, u_+(p,e')$ which is undefined
for $e' = e$. Thus, within $M^{AA}_{\mu\rho}(p,e,e')$, the fields
$A_\mu(x,e)$ and $A_\rho(x,e')$ should each have their own string
variable.

As noted in \eqref{eq:untune-that-string}, when computing the
integrands $M^{XX'}$ in propagators in $p$-space, each string
integration $I_e$ picks up an extra factor $-i u_-(p,e)$ for~$X$, and
an extra factor $i u_+(p,e') = -i u_-(p,-e')$ for~$X'$. The general
structure of the integrands is thus of the form
$$
u_-(p,e)^k u_+(p,e')^{k'} (p^2 - m^2 + i0)^{-1} M_m(p)
$$
for a suitable point-like polynomial $M_m(p)$. If this polynomial is, 
say, homogeneous of degree~$r$, the above expression will be a 
well-defined distribution on $M^4 \x H \x H$, provided that
$r - k - k' > -4$ to ensure local integrability at $p = 0$. (When
$m = 0$, the requirement is that $r - k - k' - 2 > -4$.) In the end,
products of string-like propagators are always well defined away from
$p = 0$.

\section{Debunking the axion tale}
\label{app:somnia-vana}

\begin{flushright}
\begin{minipage}[t]{22pc}
\small \itshape
Absence of evidence is not, as the saying goes, the same thing as
evidence of absence. But if you continue looking for something
intently, and keep failing to find it, you can be forgiven for
starting to worry.
\par \smallskip \hfill \upshape
--- The Economist, 12 March 2022
\end{minipage}
\end{flushright}

It should be clear that in the present formulation for QCD there is no
room for instantons and the (in)famous $\theta$-vacuum. The contention
to the purpose in SLF theory is straightforward. The string-localized
vector potentials live on Hilbert space and act cyclically on the
vacuum. In fact, every local subalgebra of operators enjoys this
property -- this is part of the Reeh--Schlieder
property~\cite{Witten18}. Therefore $\theta$-vacua are not allowed. In
plainer language: a string-localized potential for a field strength is
always given: roughly speaking, it consists of integrating the latter
along a string. So where the field-strength vanishes, the potential
remains bounded. What is more, our vector potential for gluons
possesses only the two physical degrees of freedom, so ``pure gauge''
configurations cannnot exist, then $\theta$-vacua cannot exist either,
and the so-called strong CP~problem is solved without invoking a
hypothetical particle, the \textit{axion} -- which, 46~years after its
inception, has not been found~\cite{Borisov}.

Now, if the above is the case, it should be possible, if perhaps not
quite straightfoward, to contend the same within the standard
formalism. And indeed, in~\cite{OkuboM92} it was proved by use of the
rigorous, covariant Kugo--Ojima formalism~\cite{KugoO79} that the BRS
charge ``kills'' the \textit{physical} vacuum, which if cyclic must be
unique. But that charge and the antiunitary operator for CPT
invariance commute, and this obviously demands the zero (or~$\pi$)
value for the $\theta$-parameter of the instanton makeup. One could
contend that the $\theta$-vacua are non-normalizable (a sign of
trouble in itself) and that the physical vacuum is a superposition of
them. However, by means of a simple procedure, Okubo and Marshak
showed how in that case CPT invariance still guarantees the
experimentally measurable value of~$\theta$ to be zero. We find truly
surprising in this connection that for thirty years their powerful
argument, reiterated in \cite[Chap.~10]{Marshak93}, has been
systematically ignored. In this respect, the silence of reviews of
axion theory -- like~\cite{KimC10} -- strikes us as unaccountable.

Much more recently, two important papers make the same point by a
clear, apparently different argument within the standard
approaches~\cite{AiCGT21, AiCGT22}.

It is harder still to see where the $\theta$-vacua could come from,
when using the spinor formalism -- which makes completely explicit the
counting of degrees of freedom -- to describe the gluons. In the
present respect, one still detects here the idea -- by now
discredited~\cite{Vaidman12} -- that the $A(x)$-fields ``know''
something that the $F(x)$-fields do not. It ain't so: for instance,
AQFT~\cite{LeylandRT78} tells us that Haag duality fails to hold on
causally incomplete spacetime domains, thus felicitously explaining
the Aharonov--Bohm effect without recourse to the potentials.

The moral of the story, again: quantum field theory should stand on
its own two feet, rather than on classical crutches.

\section{String-localized stress-energy-momentum tensors and dark
matter}
\label{sec:nemo}

The remorseless ascent of the ``neutrino floor'' against the searches
for dark matter (DM) by assumed ``direct detection'' particle physics
processes raises the distinct possibility that DM \textit{does not
have interactions} with the SM particles, other than through
gravitational effects.

In this respect, it has been persuasively argued in recent work by
Cheek \textit{et~al} \cite{CheekHPGT1, CheekHPGT2} that DM particles
might have been emitted as part of the Hawking radiation from
``small'' black holes emanating \textit{all} extant degrees of freedom
in nature, in the early stages of the Universe. Many current inflation
models do predict the existence of such PBH. Then dark matter would be
essentially constituted by a background of inert high-spin and
high-helicity particles $(s,|h| \geq 3)$. Those, interacting with
other particles and among themselves only via gravity, could perhaps
partially collapse into latter-day black holes (LDBH).

Whereas the elegant solution in \cite{CheekHPGT1, CheekHPGT2} to the
current predicaments appears cogent, the suggested path of DM
production in the early Universe's stages still begs the question.
This is so because the received wisdom nowadays -- well represented in
and argued~by Porrati's papers \cite{Porrati08, Porrati12} in
particular -- holds on the basis of Weinberg--Witten theorems' stiff
no-go rules \textit{against interaction with gravitons}
\cite{WeinbergW80}.

Now, precisely here SLF theory comes in succour. Perhaps the most
momentous phenomenological consequence of its use so far is that
Weinberg--Witten-like theorems are falsified: not only are
\textit{bona fide} stress-energy-momentum (SEM) tensors for massive
fields of spin $s \geq 3$ exhibited; but it is shown that for massless
fields of helicity $|h| \geq 3$ SEM tensors \textit{do exist} after
all. We quote below the explicit form of these tensors. For details,
including the continuous passage to the massless limit, we refer to
the original work \cite{MundRS17a, MundRS17b}.

\begin{itemize}
\item
For $(m > 0, s)$:
$$
T_{\al\bt}(x) = (-)^s \biggl[ -\frac{1}{4} A^\pt_{\mu\kaa}(x) 
\,\delotto_\al \,\delotto_\bt\, A^{\pt\,\mu\kaa}(x) 
- \frac{1}{2} s \,\del^\mu \Bigl( 
A^\pt_{\al\kaa}(x) \,\delotto_\bt A^{\pt\,\kaa}_\mu(x)
+ [\al \otto \bt] \Bigr) \biggr].
$$
Here the superindex~$\pt$ stands for ``Proca'' and $\kaa$ for
contraction in $s - 1$ indices $\ka_2,\dots,\ka_s$. The Proca field
$A^\pt_{\mu_1\ka_2\cdots\ka_s}$ is completely symmetric. It is
interesting that this was derived in~\cite{MundRS17b} by using SLF
theory, and restoring the above point-localized fields from their
string-localized counterparts. This is possible because $A^\pt$ enjoys
positivity.

\item
For $(m = 0, |h|)$:
\begin{align*}
T_{\al\bt}(x,e,e') 
:= \sum_{r=0}^{|h|} (-)^r & \biggl[ -\frac{1}{4} A^r_{\mu\kaa}(x,e)
\,\delotto_\al \,\delotto_\bt\, A^{r\,\mu\kaa}(x,e')
\\
&- \frac{1}{4} r \,\del^\mu \Bigl(
A^r_{\al\kaa}(x,e) \,\delotto_\bt\, A^{r\,\kaa}_\mu(x,e')
+ [e \otto e'] + [\al \otto \bt] \Bigr) \biggr].
\end{align*}
\end{itemize}

In conclusion, the essential idea in \cite{CheekHPGT1, CheekHPGT2}
gets traction from SLF theory; where the self-interaction of 
gravitons is also seen to play a role~\cite{Antigone}.

Less clearly, it seems possible that the same would apply to unbounded
helicity particles, as suggested by Schroer -- consult
\cite{Schroer17} and related earlier papers cited there. Direct
detection of~either would be well-nigh impossible, however.


\end{document}